\documentclass{iau}
\usepackage{graphicx} 
\newcommand{\bfe}{{\bf e}}
\newcommand{\bfw}{{\bf w}}
\newcommand{\ha}{\hat{\alpha}_1}
\newcommand{\haa}{\hat{\alpha}_2}

\newcommand{\wperp}{{\bf w}_\perp}
\title[IAUS291.~~New Constraints on PFEs from Binary Pulsars] 
{New Constraints on Preferred Frame Effects from Binary
  Pulsars} 
\author[L. Shao, N. Wex, \& M. Kramer]  %% short author list %%
{Lijing Shao$^1$,
 Norbert Wex$^2$,
 Michael Kramer$^3$}

\affiliation{$^1$Max-Planck-Institut f\"{u}r
  Radioastronomie, Auf dem H\"{u}gel 69, 53121 Bonn, Germany 
  \\ School of Physics, Peking University, Beijing 100871, China 
  \\ email: {\tt lshao@pku.edu.cn} 
  \\[\affilskip]
  $^2$Max-Planck-Institut f\"{u}r
  Radioastronomie, Auf dem H\"{u}gel 69, 53121 Bonn, Germany 
  \\email: {\tt wex@mpifr-bonn.mpg.de}
  \\[\affilskip]
  $^3$Max-Planck-Institut f\"{u}r
  Radioastronomie, Auf dem H\"{u}gel 69, 53121 Bonn, Germany 
  \\ Jodrell Bank Centre for Astrophysics, School of Physics
  and Astronomy,
  \\ The University of Manchester, M13 9PL, UK 
  \\ email: {\tt mkramer@mpifr-bonn.mpg.de}}

\pubyear{2012}
\volume{291}  %% insert here IAU Symposium No.
\jname{\mbox{Neutron Stars and Pulsars: 
Challenges and Opportunities after 80 years}}
\editors{J. van Leeuwen, ed.} 
\begin{document}

\maketitle

%% -- Abstract ----------------------------------
\begin{abstract}
Preferred frame effects (PFEs) are predicted by a number of
alternative gravity theories which include vector or additional tensor
fields, besides the canonical metric tensor. In the framework of
parametrized post-Newtonian (PPN) formalism, we investigate PFEs in
the orbital dynamics of binary pulsars, characterized by the two
strong-field PPN parameters, $\ha$ and $\haa$. In the limit of a small
orbital eccentricity, $\ha$ and $\haa$ contributions decouple. By
utilizing recent radio timing results and optical observations of PSRs
J1012+5307 and J1738+0333, we obtained the best limits of $\ha$ and
$\haa$ in the strong-field regime. The constraint on $\ha$ also
surpasses its counterpart in the weak-field regime.
\end{abstract}

% add below any authors, subjects and objects for indexing 
%   add more lines if necessary
%   but leave all lines commented out
%\index[author]{LastName1, Initials|textbf}
%\index[author]{LastName2, Initials|textbf}
%\index[subject]{Keyword1}
%\index[subject]{Keyword2}
%\index[object]{Object1}
%\index[object]{Object2}

\firstsection % if your document starts with a section,

\section{Introduction}

Pulsars are extremely stable electromagnetic emitters, and along with
their extreme physical properties and surrounding environments, they
provide useful astrophysical laboratories to study fundamental physics
(Lorimer \& Kramer 2005).  Radio timing of binary pulsars maps out the
binary orbital dynamics through recording the time-of-arrivals of the
pulsar signals at the telescope. For millisecond pulsars this can be
done with high precision, providing a powerful tool to probe gravity
(see e.g., \cite{sta03} and \cite{ksm+06a}).  In this work, we
summarize new results on testing the local Lorentz invariance (LLI) of
gravity from binary pulsars obtained by Shao \& Wex (2012).

%==============================================================================%

\section{New Limits on Preferred Frame Effects}

%%%%%%%%%%%%%%%%%%%%%%%%%%%%%%%%%%%%%%%%%%%%%%%%%%%%%%%%%%%%%%%%%%%%%%
% CUP work flow only accepts EPS -- not PDF, JPG, etc.
\begin{figure}[b]
\begin{center}
 \includegraphics[width=3.5in]{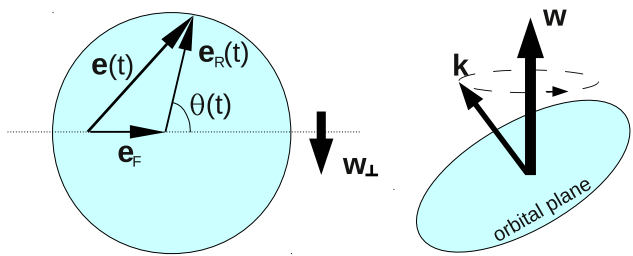} 
 \caption{Illustration of preferred frame effects in the orbital
   dynamics of small-eccentricity binary pulsars; see \cite{sw12} for
   details.  {\it Left}: $\ha$ tends to polarize the orbital
   eccentricity vector $\bfe(t)$ towards the direction perpendicular
   to $\wperp$ (Damour \& Esposito-Far{\`e}se 1992); {\it right}:
   $\haa$ induces a precession of the orbital angular momentum around
   the direction of $\bfw$.}
   \label{fig:pfe}
\end{center}
\end{figure}
%%%%%%%%%%%%%%%%%%%%%%%%%%%%%%%%%%%%%%%%%%%%%%%%%%%%%%%%%%%%%%%%%%%%%%

Non-gravitational LLI is an important ingredient of the Einstein
equivalence principle (EEP) (Will 1993, 2006). But even metric
gravity, which fulfills the EEP exactly, could still exhibit a
violation of LLI in the gravitational sector (Will \& Nordtvedt 1972,
Nordtvedt \& Will 1972, Damour \& Esposito-Far{\`e}se 1992, Will
1993). Such a violation of LLI induces preferred frame effects (PFEs)
in the orbital dynamics of a binary system that moves with respect to
the preferred frame. In the parametrized post-Newtonian (PPN)
formalism, PFEs of a semi-conservative gravity theory are described by
two parameters, $\ha$ and $\haa$.\footnote{To distinguish from their
  weak-field counterparts ($\alpha_1$ and $\alpha_2$), here ``hat''
  indicates possible modifications by strong-field effects.}

The orbital dynamics of binary pulsars with non-vanishing $\ha$ and
$\haa$ are obtained from a generic semi-conservative Lagrangian
(Damour \& Esposito-Far{\`e}se 1992). It is found that in the limit of
small orbital eccentricity, PFEs induced by $\ha$ and $\haa$ decouple,
and lead to separable effects in the timing observations. Hence they
can be tested independently using observations of only one binary
pulsar (Shao \& Wex 2012).

\begin{itemize}
\item A non-zero $\ha$ induces a polarization of the eccentricity
  vector towards a direction in the orbital plane perpendicular to the
  velocity of the binary system with respect to the preferred frame,
  $\bfw$.\footnote{Here we choose the isotropic cosmic microwave
    background as the preferred frame; nevertheless, see \cite{sw12}
    for constraints on other preferred frames.}  The observed
  eccentricity vector $\bfe(t)$ is a vectorial superposition of a
  ``relativistically rotating'' eccentricity $\bfe_R(t)$ (of constant
  length) and a ``fixed eccentricity'' $\bfe_F \propto \ha$: $\bfe(t)
  = \bfe_R(t) + \bfe_F$ (Damour \& Esposito-Far{\`e}se 1992). The
    effect is graphically illustrated in the left panel of
  Fig.~\ref{fig:pfe}. Previous methods use the smallness of the
  observed eccentricity, combined with probabilistic considerations
  concerning the unknown angle $\theta$ (the angle between $\bfe_R$
  and $\bfe_F$) to constrain $\ha$ (Damour \& Esposito-Far{\`e}se
  1992, Wex 2000). The method developed in \cite{sw12} is an extension
  of the method by Damour \& Esposito-Far{\`e}se (1992) that does not
  require any probabilistic considerations concerning $\theta$. It is
  applicable to binary pulsars of short orbital period that have been
  observed for a long enough time, during which the periastron has
  advanced significantly. Even if the advance of periastron is not
  resolved in the timing observation, the constraints on the
  (observed) eccentricity vector can be converted into a limit on
  $\ha$. From the 10 years of timing and the optical observations of
  PSR~J1738+0333 (Antoniadis \etal\ 2012, Freire \etal\ 2012) one
  obtains the most constraining limit,
  \begin{equation}\label{eq:a1}
    \ha = - 0.4^{+3.7}_{-3.1} \times 10^{-5} \quad \mbox{(95\% C.L.)} \,,
  \end{equation}
  which is significantly better than the current best limits from
  both weak field (M{\"u}ller \etal\ 2008) and strong field (Wex
  2000).

\item A non-vanishing $\haa$ induces a precession of the orbital angular
  momentum around the direction of $\bfw$ (see the right panel of
  Fig.~\ref{fig:pfe}), which changes the observed orbital inclination. 
  Using the long-term timing results on PSRs J1012+5307 (Lazaridis \etal\ 2009) 
  and J1738+0333 (Antoniadis \etal\ 2012, Freire
  \etal\ 2012), \cite{sw12} find an upper limit
  \begin{equation}\label{eq:a2}
    |\haa| < 1.8 \times 10^{-4} \quad \mbox{(95\% C.L.)} \,,
  \end{equation}
  which is better than the current best limit for strongly self-gravitating
  bodies (Wex \& Kramer 2007) by more than three orders of magnitude, but still 
  considerably weaker than the weak-field limit of $\alpha_2$ by Nordtvedt 
  (1987).
\end{itemize}

%==============================================================================%

\section{Summary}

We summarize results presented in Shao \& Wex (2012) that proposed new
tests of LLI. These yield improved constraints on PFEs from binary
pulsar experiments. Specifically, limits on parameters $\ha$ and
$\haa$ are obtained from long-term timing of two binary pulsars with
short orbital period (see Eqs.  (\ref{eq:a1}) and (\ref{eq:a2})).  Our
extended $\ha$ test no longer requires probabilistic considerations
related to unknown angles. The proposed tests have the advantage that
they continuously improve with time, and will benefit greatly from the
next generation of radio telescopes, like FAST (Nan \etal\ 2011) and
SKA (Smits \etal\ 2009).

%============================================================

\section*{Acknowledgment}

Lijing Shao is supported by China Scholarship Council (CSC).

%==============================================================================%

\end{document}